\documentclass{osa-article}
\usepackage[subpreambles=true]{standalone}
\usepackage{gensymb}
\usepackage{xcolor}
\usepackage{multirow}
\usepackage{subfiles}
\journal{optica}
\articletype{Research Article}
\usepackage{chapterbib}

\begin{document}

%\documentclass{osa-article}
%\usepackage{gensymb}
%\usepackage{xcolor}
%\usepackage{multirow}
%\usepackage{subfiles}
%\journal{optica}
%\articletype{Research Article}
%\begin{document}

\title{Reflection confocal nanoscopy using a super-oscillatory lens}
\author{Arvind Nagarajan\authormark{1,2,*}, L. Pjotr Stoevelaar\authormark{1,2}, Fabrizio Silvestri\authormark{2}, Marijn Siemons\authormark{3,4}, Venu Gopal Achanta\authormark{5}, Stefan M. B.  B\"{a}umer\authormark{2}, and Giampiero Gerini\authormark{1,2}}

\address
{\authormark{1}Electromagnetics Group, Technische Universiteit Eindhoven (TU/e), 5600 MB Eindhoven, The Netherlands\\
\authormark{2}Optics Department, Netherlands Organization for Applied Scientific Research (TNO),\newline Stieltjesweg 1, 2628 CK Delft, The Netherlands\\
\authormark{3}Department of Imaging Physics, Delft University of Technology, 2628 CJ Delft, The Netherlands\\
\authormark{4}Currently with Cell Biology, Department of Biology, Faculty of Science, Utrecht University, 3584 CH Utrecht, The Netherlands\\
\authormark{5}FOTON Laboratory, Tata Institute of Fundamental Research, Homi Bhabha Road, Mumbai 400005, India\\}

\email{\authormark{*}arvind.nagarajan@tno.nl} 

%%%%%%%%%%%%%%%%%%% abstract %%%%%%%%%%%%%%%%
\begin{abstract*}
A Superoscillatory lens (SOL) is known to produce a sub-diffraction hotspot which is useful for high-resolution imaging. However, high-energy rings called sidelobes coexist with the central hotspot. Additionally, SOLs have not yet been directly used to image reflective objects due to low efficiency and poor imaging properties. We propose a novel reflection confocal nanoscope which mitigates these issues by relaying the SOL intensity pattern onto the object and use conventional optics for detection. We experimentally demonstrate super-resolution by imaging double bars with 330 nm separation using a 632.8 nm excitation and a 0.95 NA objective. We also discuss the enhanced contrast properties of the SOL nanoscope against a laser confocal microscope, and the degradation of performance while imaging large objects.
\end{abstract*}
%%%%%%%%%%%%%%%%%%%%%%%%%%  body  %%%%%%%%%%%%%%%%%%%%%%%%%%
\section{Introduction}

Optical microscopy is an essential imaging technique with a wide range of applications, such as, microelectronics, mineralogy, and microbiology \cite{Herman1993}. Its fundamental limitation is the restricted spatial resolution due to diffraction \cite{Rayleigh1903}. Abb{\'e}, for example, uses the criterion  $\Delta x=\lambda/(2NA)$ to determine the limit $\Delta x$ in the spatial resolution while illuminating with light at a wavelength $\lambda$ \cite{Abbe1873}. Here, NA is the effective numerical aperture of the imaging system. Consequently, great attempts have been made in the past century to enhance the spatial resolution of optical microscopy by either decreasing the operational wavelength or by increasing the NA. The resulting techniques include a plethora of near-field solutions \cite{Pendry2000,Jacob2006,Mansfield1990, Betzig1991,Wang2011,Yan2014} which achieve super-resolution by capturing evanescent waves, thereby increasing the NA. The operational wavelength has also been reduced to DUV \cite{Vollrath2005} and EUV \cite{Wachulak2010} to enhance the spatial resolution. Alternatively, there are also fluorescence based techniques \cite{Hell1994,Betzig2006,Gustafsson2000} which rely on quantum emitters to achieve super-resolution. \par   

Although many of these techniques achieve a spatial resolution ranging from several tens to a hundred nanometres, they compromise on several advantages of an optical microscope. Optical superoscillation is a promising technique to achieve super-resolution in the far-field without the need to capture the evanescent waves or decrease the operational wavelength.  It can produce a focal spot much smaller than the Abb{\'e} diffraction limit by carefully tailoring the interference of a large number of beams diffracted from a nanostructured mask \cite{Rogers2012a}. Thus far, this technique has only been employed in transmission \cite{Gbur2019}. It would be favorable to develop a non-invasive, far-field, super-resolution reflection microscope (or, more precisely, nanoscope) to image non-transmissive objects for applications such as semiconductor metrology \cite{Shankar2005} or imaging OLEDs \cite{King2016}, among others. \par

The concept of superoscillation was first introduced in 1952 by Toraldo di Francia \cite{DiFrancia1952} by proposing the idea of a super-gain antenna to improve the imaging resolution. More recently, Berry and Popescu \cite{Berry2006} found that superoscillation behaviour occurs when a waveform appears to locally oscillate faster than its highest spectral component, analogous to weak measurements in quantum mechanics \cite{Aharonov1988}. There are few well-established methods to design a SOL at optical frequencies: using a binary metal mask \cite{Rogers2012a}, a phase mask \cite{Wong2013}, or by using radial-polarization \cite{Kozawa2018}. These can all produce a central "hotspot" which is much smaller than the Abb{\'e} diffraction limit. Also, broadband achromatic SOLs have been experimentally demonstrated \cite{Tang2015,Li2018}. However, it is not straightforward to use the SOL in reflection mode for simultaneous illumination and pickup due to issues such as high intensity background (from reflective binary metal masks in SOLs \cite{Ferreira2006}) or poor imaging properties due to aberrations (in phase mask SOLs). Yet, by combining SOL illumination with a diffraction-limited high NA lens the advantages of super-oscillation can be retained as demonstrated in this work. \par

Additionally, a major obstacle with the SOL is the existence of a high-intensity region (called sidelobes) close to the central sub-diffraction hotspot (see insets of Fig.\ref{fig:setup}\textcolor{urlblue}{(b)}). Field of View (FOV), a metric quantifying the sidelobes, is defined as the region within the first sidelobe containing the central hotspot \cite{Rogers2012a}. These sidelobes are intense compared to the central hotspot (see \textcolor{urlblue}{Supplementary Figure S4, S5}), and their intensity is inversely related to the size of the hotspot. Although these sidelobes cannot be completely eliminated (without destroying the superoscillaory feature), they can be pushed away to have a bigger FOV \cite{Rogers2013,Lindberg2012,Kempf2018,Hyvarinen2012,Rogers2018} or their intensities can be reduced \cite{Qin2017, Rogers2013a,Yuan2015,Diao2016, Dong2017}, both at the cost of an increased central hotspot size. In the case of extended objects (objects larger than the FOV), these sidelobes illuminate regions of the object off-axis. The reflected/transmitted signal from these regions can reach the detector on-axis potentially degrading the imaging properties. An investigation of the consequences of sidelobe illumination in imaging extended objects is necessary and is addressed in this work. All previous works on SOLs have only demonstrated super-resolution capabilities in transmission mode by imaging objects within the FOV, and imaging extended objects were limited to an ensemble of isolated objects.  \par

In this paper, we report the first experimental demonstration of a reflection confocal nanoscope employing a binary metal mask SOL. The SOL is first characterized (in reflection) over a range of distances, and a suitable sub-diffraction limited hotspot is chosen to image a series of objects. The objects, consisting of double-bars, 1D and 2D arrays are designed to determine the influence of sidebands in the imaging capabilities. The results show that although the SOL can image objects which fit within the FOV with super-resolution, it has poor imaging properties while imaging extended objects.\par

\section{Materials and Methods}

A SOL used for pickup suffers from aberrations, thereby limiting its use as an illumination lens only. This concept has been previously used in a transmission confocal setup \cite{Rogers2012a}. Yet, an equivalent reflection confocal setup can be conceived as shown in Fig.\ref{fig:setup}\textcolor{urlblue}{(a)}. This setup, elaborated in the following subsection, combines the SOL illumination and the confocal reflection mode advantages by relaying the SOL intensity pattern onto the object and exploits the polarization to isolate the reflected beam from the illumination beam. \par  

\subsection{Experimental setup}

\begin{figure}[htbp]
\centering\includegraphics[width=13cm]{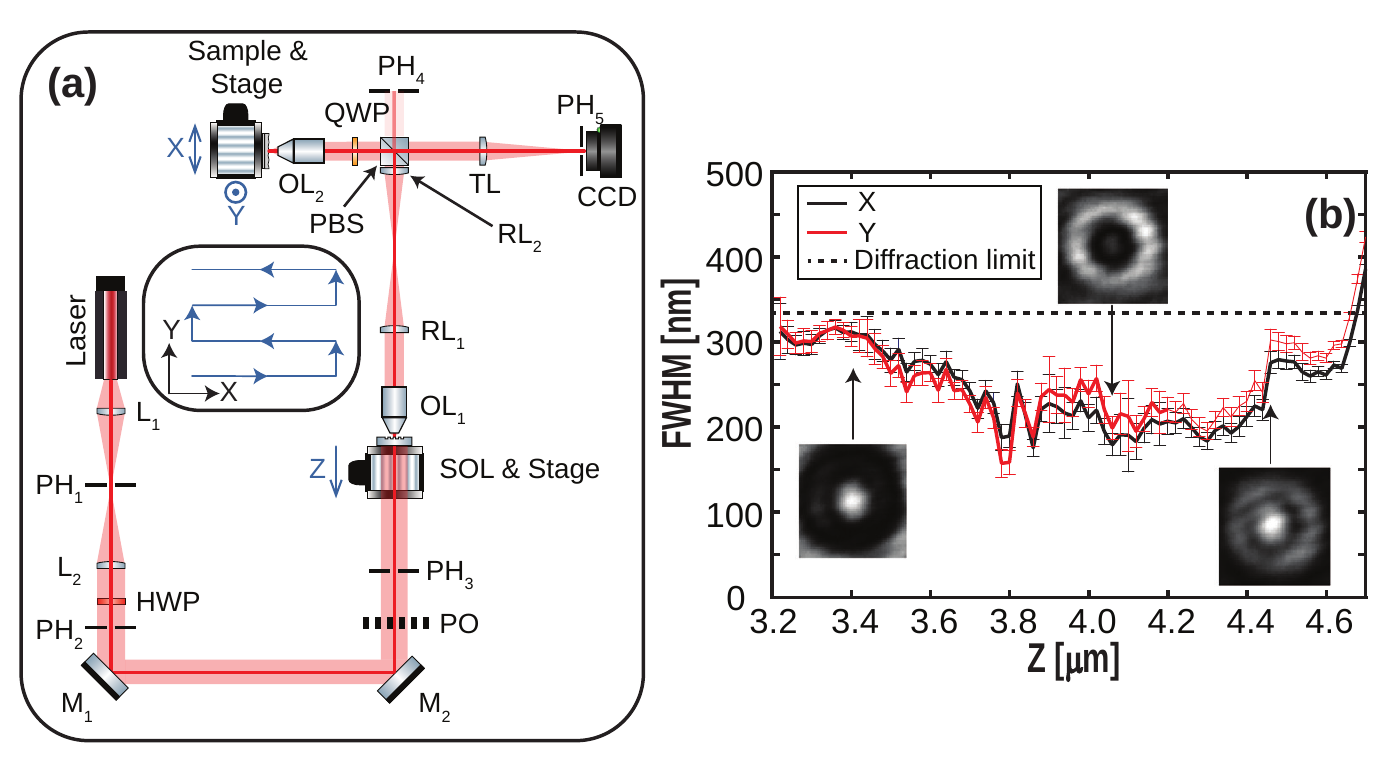}
\caption{\textbf{Experimental Setup and SOL Characterization.} \textbf{(a)} Schematic of the experimental setup. The SOL is illuminated by a 632.8 nm collimated polarised laser beam. The light intensity pattern produced by the SOL is reimaged onto the sample plane using a pupil relay. The object is snake scanned in steps of 30 nm in the XY plane (see inset), and the reflected signal is focused into the CCD with a tube lens. \textbf{(b)} Full width at half maximum (FWHM) of the central hotspot for various Z distances from the SOL surface. The red and black curves indicate the FWHM across the X and Y planes respectively. The error bars indicate 95$\%$ confidence interval in the gaussian fitting. The black dotted line represents the Abb{\'e} diffraction limit (333 nm). Insets show the intensity-patterns at various Z distances captured by the CCD.}
\label{fig:setup}
\end{figure}

We built a modified laser scanning confocal microscope (LSCM) setup with SOL illumination to realize super-resolution imaging in reflection mode, as illustrated in Fig.\ref{fig:setup}\textcolor{urlblue}{(a)}. The intensity pattern produced by the SOL from a 632.8 nm polarised laser is captured by the first objective lens $(OL_{1})$ (Nikon M Plan APO 150X, 0.95 NA) and relayed by two lenses $(RL_{1}, RL_{2})$ (f = 100 mm) to the second objective lens $(OL_{2})$ (Leica HCX PL APO 150X, 0.95 NA) resulting in a far-field super-resolution illumination system. The objectives should have an adequately high NA so that all  the spatial frequency content necessary for superoscillation is picked up (see \textcolor{urlblue}{Supplement} for elaboration). The pupil relay is essential to effectively transfer this high-frequency content from $OL_{1}$ to  $OL_{2}$ thereby preserving the superoscillation. All components of the pupil relay ($OL_{1}$, $OL_{2}$, $RL_{1}$ and $RL_{2}$) must be carefully aligned to achieve the spatial frequency preserving relay. \par

The polarizing beam splitter (PBS) and the quarter wave-plate (QWP) isolates the illumination and the detection beams. The reflected beam picked up by $OL_{2}$ is transmitted at the PBS and is subsequently focused by a tube lens (f = 200 mm) on a Charge-coupled device (CCD) camera (Thorlabs DCU223M, pixel size = $4.65 \times 4.65$ $\mu m^2$, $1024 \times 768$ pixels). \par 
The SOL and object are mounted on piezo controlled stages (Thorlabs - MAX311D/M, closed loop). The piezo stages are in turn mounted on a high load pitch and yaw platform (Thorlabs - PY004/M) to control the direction of the stage movement. The pitch and yaw of both stages are corrected to <2 mrad precision to allow a maximum defocus of 20 nm in the Z direction while scanning a 10 $\mu$m area in XY direction to minimize aberrations. \par

The setup described here can be easily modified to do LSCM imaging (reported in Fig.\ref{fig:bars}.) by removing the SOL and the $OL_{1}$ from the illumination beam path. \par

\subsection{SOL characterization}

We characterize the hotspot of the manufactured SOL (see \textcolor{urlblue}{Supplement} for device fabrication) over a distance to account for a small change in the excitation wavelength and artefacts in the device fabrication \cite{Yu2018}.  The SOL intensity pattern is imaged onto a reflective gold alignment pad ($100 \mu m \times 20 \mu$m) in the object. The detection chain picks up the reflected signal, and we characterize the intensity (I (x, y)) from the CCD images. As the SOL field pattern goes through the entire optical chain for characterization, we are also examining if the pupil relay adequately transfers the high-frequency content from $OL_{1}$ to $OL_{2}$. \par
We scan the SOL in the Z direction (see Fig.\ref{fig:setup}\textcolor{urlblue}{(a)}) from 0 (SOL surface) to 20 $\mu$m in steps of 20 nm to characterize its depth of focus (DOF) along with the position of a sub-diffracted hotspot with a large FOV. The central hotspot is predominantly circular in nature. A least-square fit with a 2D Gaussian as shown in Eq.(\ref{eqn_1}) can be applied to it as it readily gives the FWHM.\par
\begin{equation}\label{eqn_1}
    I(x,y)=I_{max}\exp\bigg(\frac{-(x-x_{0})^2}{2\sigma_{x}^2}+\frac{-(y-y_{0})^2}{2\sigma_{y}^2}\bigg)+I_{off}
\end{equation}
Here, $I_{max}$ is the peak intensity of the hotspot, $(x_{0},y_{0})$ is the position of the hotspot centre, $\sigma_{x}$ and $\sigma_{y}$ are the 1$/(2e^2)$ points along x and y and $I_{off}$ is the offset due to background signal.  We calculate the FWHM from the beam waist using Eq.(\ref{eqn_2}) and Fig.\ref{fig:setup}\textcolor{urlblue}{(b)} displays the FWHM for a region with super-resolution.\par
\begin{equation}\label{eqn_2}
    FWHM=\sigma\Big[2\sqrt{2\ln(2)}\Big]
\end{equation}
The red and black curves in Fig.\ref{fig:setup}\textcolor{urlblue}{(b)} represent the FWHM in X and Y directions respectively, and the error bars indicate the $95\%$ confidence interval of the least-squares fitting. The black dotted curve represents the Abb{\'e} diffraction limit for a LSCM($\approx 333 nm$). The sub-diffraction hotspot is predominantly circular in nature as the two curves almost overlap with each other within the $95\%$ confidence intervals of the fitting and has the potential to substantially enhance spatial resolution compared to a LSCM. Some of the raw images captured by the CCD camera for various Z distances are shown in the insets. The FWHM changes gradually and varies by $<20$ nm in a 100 nm range, apart from few outliers where the change is $>100$ nm. The DOF of the SOL is hence deduced to be $\approx$100 nm. The hotspot around $Z=4\mu$m has a FWHM of $189 \pm 21$ nm, an effective numerical aperture ($NA_{eff}$)  of $1.674 \pm 0.19$ (using the definition of $FWHM=\lambda/(2NA_{eff})$ \cite{Ni2018}) and has the largest field of view (FOV) of $\approx$500 nm. We fix the SOL at this distance and all images/simulations reported in this paper are with this intensity pattern of the SOL. \par

\subsection{Simulation of imaging with the SOL nanoscope}

We simulate the point-scanning confocal imaging reported in the next section using linear systems theory assuming a coherent source\cite{Goodman1996}. A 2D convolution of the Coherent spread function (CSF) of both illumination ($CSF_{I}$) and detection ($CSF_{D}$), with the object O(x, y), gives its image G($\epsilon, \phi$) as specified in Eq.(\ref{eqn_3}). The experimentally obtained point spread function (PSF) of the SOL is used to derive the illumination CSF mentioned in Eq.(\ref{eqn_4}) as the phase is not experimentally measured. This might cause some discrepancy between simulations and measurements, nevertheless giving a first approximation of the expected image. The detection CSF is a jinc function given by Eq.(\ref{eqn_4}), where $J_{1}$ is the Bessel function of the first kind of order 1, and $\lambda$ is the operating wavelength (632.8 nm). Also, the object is modelled as a binary function with $100\%$ reflection from the structure and $0\%$ reflection from background ignoring the thickness. This procedure results in a small intensity discrepancy between the simulations and the measurements. \par

\begin{equation}\label{eqn_3}
    G(\epsilon,\phi)=\left|\left|\int_{y_{min}}^{y_{max}}\int_{x_{min}}^{x_{max}} {CSF_{I}(x-\epsilon,y-\phi)}CSF_{D}(x-\epsilon,y-\phi)O(x,y)dxdy\right|\right|^2
\end{equation}
\begin{equation}\label{eqn_4}
    CSF_{I}(x,y)\approx\sqrt{PSF_{SOL}(x,y)} \; \; \; \;\&\; \; \; \;
    CSF_{D}(x,y)=\frac{2J_{1}\bigg(\frac{2\pi NA}{\lambda}\sqrt{x^2+y^2}\bigg)}{\bigg(\frac{2\pi NA}{\lambda}\sqrt{x^2+y^2}\bigg)}
\end{equation}

\subsection{Imaging under SOL illumination}

The test structures are point-scanned with a step size of 30 nm (see \textcolor{urlblue}{Supplement} for elaboration) in both X and Y directions in a snake like pattern as illustrated in the inset of Fig.\ref{fig:setup}\textcolor{urlblue}{(a)}. The reflected signal is captured by the CCD camera, and as suggested in \cite{Rogers2012a}, a confocal pinhole $(PH_{5})$ is numerically implemented by recording the average pixel intensity corresponding to the central region ($40\%$ of the hotspot diameter, $3 \times 3$-pixel grid centred at $(x_{0}, y_{0})$) of the SOL hotspot for each scan position. Hence, the image is reconstructed without any deconvolution or post-processing. \par

\section{Results and discussions}

We image the following three classes of test structures (see \textcolor{urlblue}{Supplement} for device fabrication) with characteristic dimensions both above and below the Abb{\'e} diffraction limit ($\approx$333  nm) using a super-oscillatory hotspot with a FWHM of $189 \pm 21$ nm and a FOV of $\approx$500 nm using an excitation wavelength of 632.8 nm and a $150X$, 0.95 NA focusing$/$collecting objective lens (see \textcolor{urlblue}{Supplementary Figure S4, S5}). \par

\begin{enumerate}
\item \textbf{Objects which fit within the FOV of the SOL:} Two sets of double bars (500 nm $\times$ 180 nm) with a centre to centre (c.t.c.) spacing of 500 nm and 330 nm respectively. 
\item \textbf{1D array:} Arrays consisting of 10 bars (500 nm $\times$ 180 nm) with a c.t.c. spacing of 500 nm and 330 nm respectively.
\item \textbf{2D objects:} (a) Cluster of circles of diameter 200 nm with a c.t.c. spacing ranging from 240 nm to 480 nm. (b) A $10 \times 10$ array of squares of size 100 nm in a square lattice with 280 nm periodicity. \par
\end{enumerate}

It is to be noted here that the conventional rules for two-point resolution such as the Rayleigh criterion \cite{Rayleigh1903} do not apply for SOL imaging as the location of the first zeros do not coincide. Hence we define and follow a simple rule for SOL imaging, which is a variant of the Rayleigh limit and applies for any super-resolution lens. A peak-valley-peak $(P_{1}-V-P_{2})$ system is resolved when the difference between the measured peak and valley intensities (d.p.v) is at least $10\%$ as given by Eq.(\ref{eqn_5}). Here $I_{P_{1}}, I_{P_{2}}$ and $I_{V}$ are normalized peak and valley intensities respectively. \par

\begin{equation}\label{eqn_5}
    d.p.v=min (I_{P_{1}},I_{P_{2}})-I_{V}
\end{equation}

\begin{figure}[htbp]
\centering\includegraphics[width=13cm]{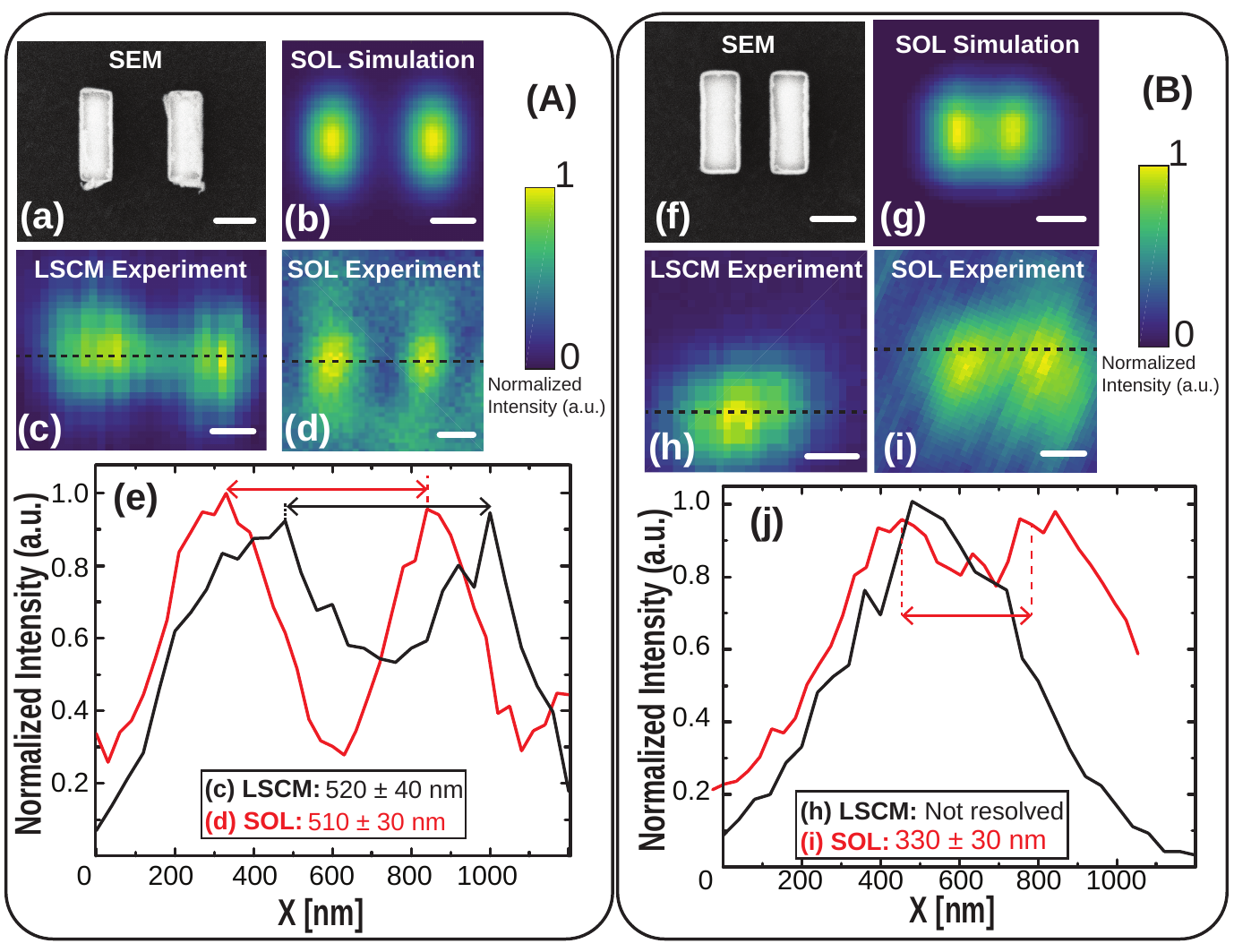}
\caption{\textbf{SOL nanoscope vs LSCM: Imaging double bars.} \newline Double bars (500 nm $\times$ 180 nm) with c.t.c. separation of \textbf{(A)} 500 nm and \textbf{(B)} 330 nm respectively. \textbf{(a, f)} SEM images of Au bars on ITO glass substrate. \textbf{(b, g)} Numerical simulation of imaging with the SOL nanoscope. \textbf{(c, h)} LSCM imaging. \textbf{(d, i)} Imaging with the SOL nanoscope. \textbf{(e, j)} Intensity profiles along the central black dashed lines in (c, d) and (h, i) respectively. The white scale bar represents 200 nm.}
\label{fig:bars}
\end{figure}

Fig.\ref{fig:bars}. shows a comparison of imaging with the SOL nanoscope and a conventional LSCM on resolving the double bars with a c.t.c. spacing above (Panel A) and below (Panel B) the Abb{\'e} diffraction limit. Although both the LSCM and the SOL nanoscope could resolve the double bars shown in Fig.\ref{fig:bars}\textcolor{urlblue}{(a)}, they appear clearer with the SOL nanoscope. As inferred from the intensity profiles in Fig.\ref{fig:bars}\textcolor{urlblue}{(e)}, the LSCM reports a c.t.c. separation of 520 $\pm$ 40 nm with $41.49\%$ d.p.v.  The SOL nanoscope reports a c.t.c. separation of 510 $\pm$ 30 nm, with a $72.3\%$ d.p.v. The experimentally obtained image, although noisy, agrees quite well with the numerical simulations shown in Fig.\ref{fig:bars}\textcolor{urlblue}{(b)} which predicts a c.t.c. of 500 nm with a d.p.v of $86.4\%$. As expected, the LSCM could not resolve the double bars with c.t.c. separation of 330 nm shown in Fig.\ref{fig:bars}\textcolor{urlblue}{(f)}. The SOL nanoscope, on the other hand, could resolve them. Numerical simulations shown in  Fig.\ref{fig:bars}\textcolor{urlblue}{(g)} predicts a c.t.c. of 330 nm with a $20\%$ d.p.v. Experimentally a c.t.c. separation of 330 $\pm$ 30 nm with $15.5\%$ d.p.v. is observed from the line profile shown in Fig.\ref{fig:bars}\textcolor{urlblue}{(j)}. This demonstrates the super-resolution capability of the SOL nanoscope in imaging reflective objects which fit within the FOV. \par

\begin{figure}[htbp!]
\centering\includegraphics[width=13cm]{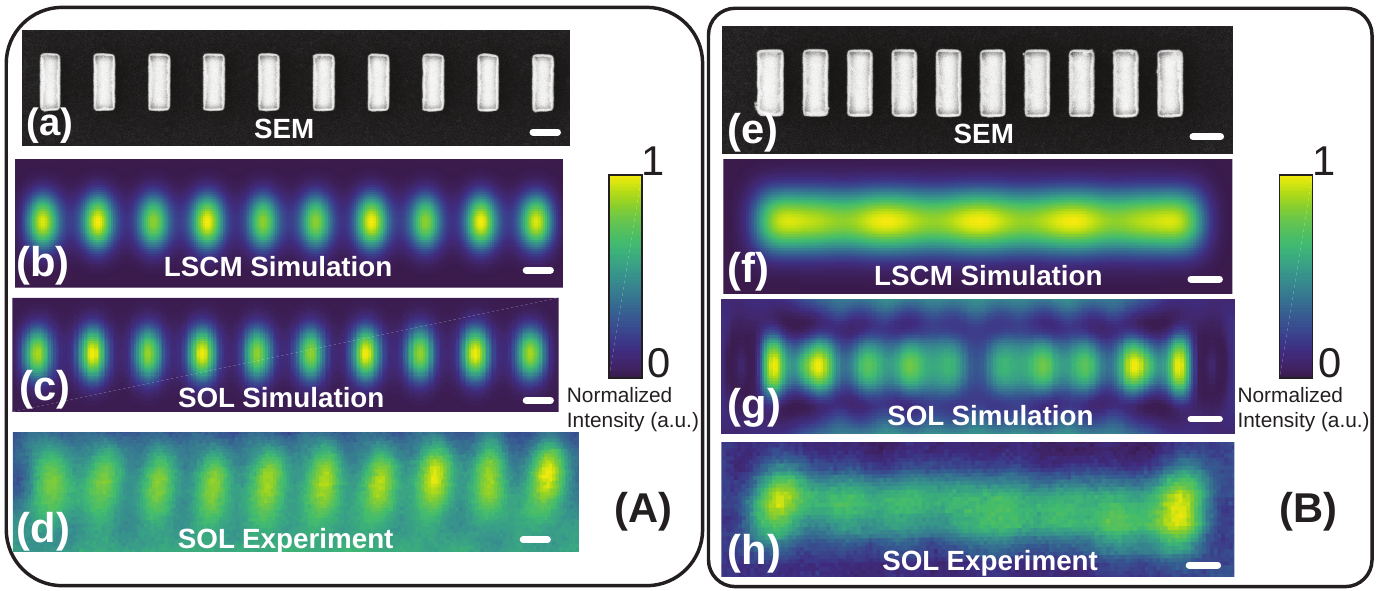}
\caption{\textbf{Imaging 1D array with a c.t.c. spacing (A)bove \& (B)elow the Abb{\'e} diffraction limit}. Arrays consisting of 10 bars (500 nm $\times$ 180 nm) with c.t.c. separation of \textbf{(A)} 500 nm and \textbf{(B)} 330 nm respectively. \textbf{(a, e)} SEM images of Au arrays on ITO glass substrate. \textbf{(b, f)} Numerical simulation of imaging with a LSCM. \textbf{(c, g)} Numerical simulation of imaging with the SOL nanoscope. \textbf{(d, h)} Imaging with the SOL nanoscope. The white scale bar represents 200 nm.}
\label{fig:1D}
\end{figure}

The double bars structure is extended to form a 1D array consisting of 10 bars to investigate the impact of sidelobe illumination.  Fig.\ref{fig:1D}  summarizes the performance of both the LSCM and the SOL nanoscope in imaging these 1D arrays with the c.t.c. spacing above (panel A) and below (panel B) the Abb{\'e} diffraction limit. Although numerical simulations indicate that both the LSCM (Fig.\ref{fig:1D}\textcolor{urlblue}{(b)}) and the SOL nanoscope (Fig.\ref{fig:1D}\textcolor{urlblue}{(c)}) could resolve the 1D array with a c.t.c. spacing of 500 nm, the contrast is better with the SOL nanoscope. An average d.p.v of $68.35\%$ is expected from the LSCM, while the SOL nanoscope simulations suggest an average d.p.v of $78.11\%$ (see \textcolor{urlblue}{Supplementary Figure S7} for line profile). The experimental results shown in Fig.\ref{fig:1D}\textcolor{urlblue}{(d)} are noisy and report a c.t.c. separation of 510 $\pm$ 30 nm with an average d.p.v. of $34.42\%$ (see \textcolor{urlblue}{Supplementary Figure S8} for line profile). The discrepancy between the numerical simulations (Fig.\ref{fig:1D}\textcolor{urlblue}{(c)}) and the experimental results (Fig.\ref{fig:1D}\textcolor{urlblue}{(d)}) is attributed to the neglection of the actual phase of the sidelobes in the measured PSF used in the simulations. (see \textcolor{urlblue}{Supplement} for elaboration). Also, the substantial dip in the experimentally measured d.p.v (Fig.\ref{fig:1D}\textcolor{urlblue}{(d)}) compared to the case of the double bar (Fig.\ref{fig:bars}\textcolor{urlblue}{(d)}) is deduced to be due to the sidelobe illumination as it is the only difference between the two systems. When we further reduce the c.t.c. spacing of the array to 330 nm, the influence of sidelobe illumination becomes unambiguous as seen in Panel (B) of Fig.\ref{fig:1D}. While the LSCM clearly fails to resolve this array (Fig.\ref{fig:1D}\textcolor{urlblue}{(f)}), numerical simulations on the SOL nanoscope (Fig.\ref{fig:1D}\textcolor{urlblue}{(g)}) predict a weak ringing effect and even though 10 bright spots (corresponding to 10 bars) are seen, they appear to get closer towards the center of the array. The peak intensities corresponding to the bars is not uniform across the array and it drops by $36\%$ at the center of the array with respect to the edges. Bars $4-5$ and $6-7$ are almost merged with a d.p.v. of $6\%$, while the remaining bars are resolved with d.p.v. $>18\%$. Experimentally (Fig.\ref{fig:1D}\textcolor{urlblue}{(f)}.), only the first and the last bars are noticeably intense with a d.p.v. of $30\%$, and all other inner bars are not resolved (d.p.v $<10\%$) (see \textcolor{urlblue}{Supplementary Figure S9} for line profile). Table \ref{tab:dpv} summarises the d.p.v. for both the LSCM and the SOL nanoscope in imaging double bars and 1D arrays.  \par

\begin{figure}[htbp]
\centering\includegraphics[width=5cm]{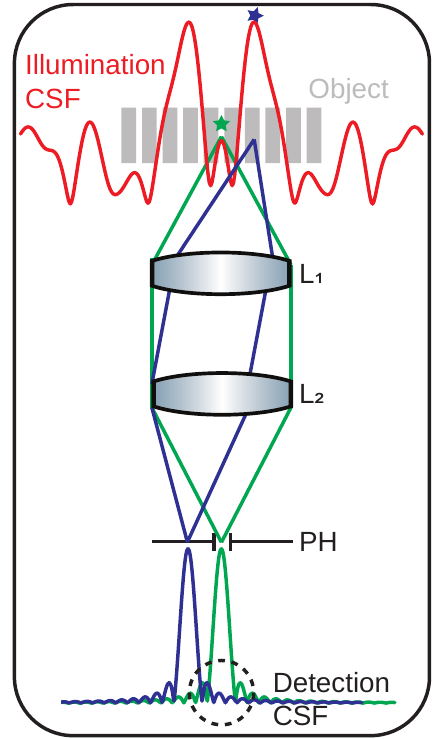}
\caption{\textbf{Leakage into the detector due to sidelobe illumination}.\newline
A closely spaced extended object (gray) is imaged with a SOL CSF ($CSF_{I}$) (red curve) using a lens system ($L_{1}, L_{2}$). Although the rays from the sidelobe illumination (blue) do not seem to enter the pinhole (PH), there is a noticeable leakage from the ripples of the detection CSF ($CSF_{D}$) at the detector as highlighted by the black dotted circle. The green rays show the imaging from the central hotspot. The stars indicate the intensities of the central hotspot and the first sidelobe, respectively.}
\label{fig:sidelobe}
\end{figure}

Although we here only image two sets of isolated objects and 1D arrays, they are representative of the influence of sidelobe illumination. If an ensemble of isolated objects are separated by a distance much larger than the FOV, there is a reduced information cross-coupling between the objects, and each of them can be imaged with high-resolution. However, when these objects are close together, the leakage of the detection CSF, combined with the higher intensity of the illumination CSF for an off-axis point, will determine a background signal higher than the required on-axis one as explained in Fig.\ref{fig:sidelobe}.  

\begin{table}[htbp!]
\centering
\caption{d.p.v. of the SOL nanoscope compared to a LSCM.}
\begin{tabular}{|c|c|c|c|c|}
    \hline
    \bf Structure & \bf c.t.c. (nm) & \bf LSCM & \bf SOL simulation & \bf SOL experiment\\
    \hline
    \multirow{2}{4em}{\bf Double bars} & 500 & 41.49$\%$ (experiment) & 86.4$\%$ & 72.3$\%$ \\\cline{2-5}
    & 330 & 0$\%$ (experiment) & 20$\%$ & 15.5$\%$\\
    \hline
    \multirow{2}{4em}{\bf 1D array} & 500 & 68.35$\%$ (simulation) & 78.11$\%$ & 34.42$\%$\\\cline{2-5}
    & 330 & 0$\%$ (simulation) & <10$\%$ & <10$\%$ \\
    \hline
\end{tabular}
\label{tab:dpv}
\end{table}

\begin{figure}[htbp]
\centering\includegraphics[width=13cm]{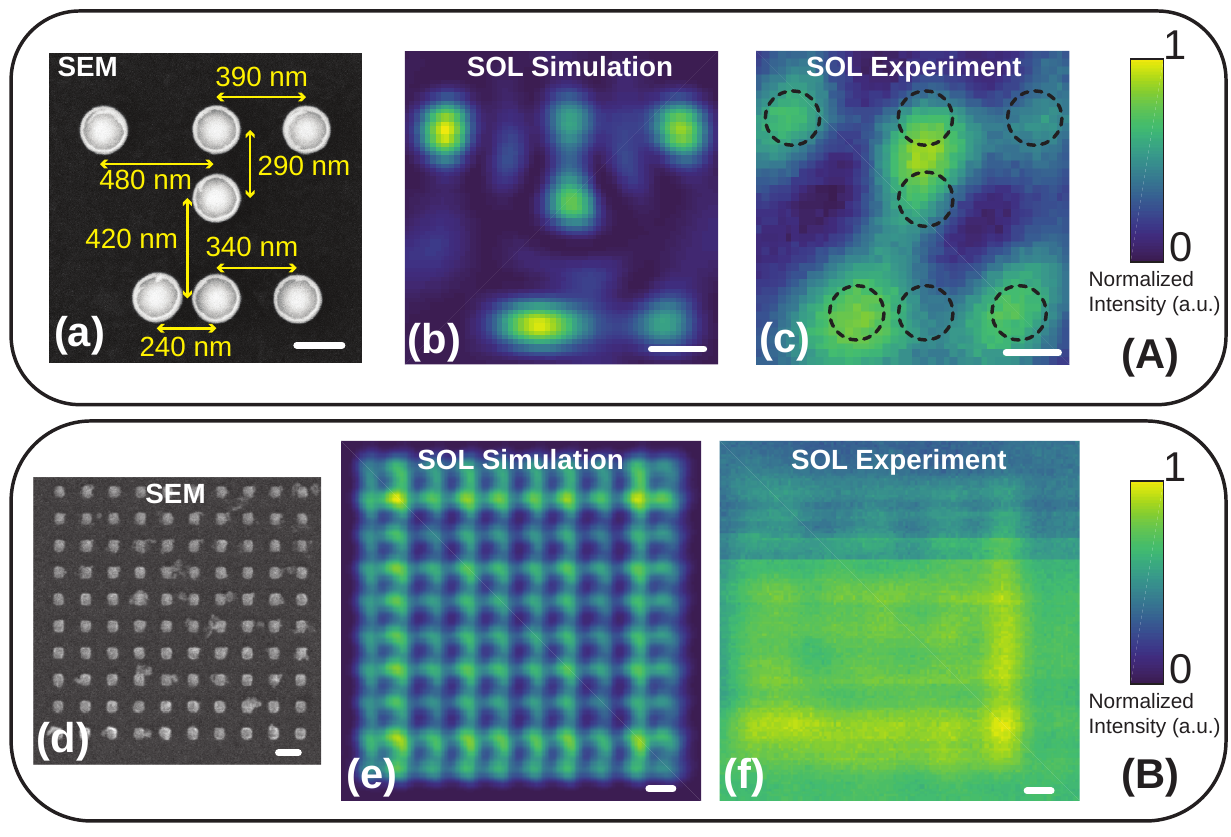}
\caption{\textbf{Imaging 2D objects under SOL illumination}\newline
\textbf{(A)}bove: Cluster of circles of diameter 200 nm with various c.t.c. separations.\newline
\textbf{(B)}elow: $10 \times 10$ array of squares of size 100 nm in a square lattice with 280 nm periodicity. \textbf{(a, d)} SEM images of Au arrays on ITO glass substrate. \textbf{(b, e)} Numerical simulation of imaging under SOL illumination. \textbf{(c, f)} Imaging with the SOL nanoscope. The white scale bar represents 200 nm.}
\label{fig:2d}
\end{figure}

We further increase the complexity of the test structures by imaging two classes of 2D objects with the SOL nanoscope. It consists of an ensemble of circles of diameter 200 nm with various c.t.c. separations as seen in Fig.\ref{fig:2d}\textcolor{urlblue}{(a)}. and a 10$\times$10 array of squares of size 100 nm in a square lattice with periodicity of 280 nm as seen in Fig.\ref{fig:2d}\textcolor{urlblue}{(d)}. Numerical simulations (Fig.\ref{fig:2d}\textcolor{urlblue}{(b)}) predict most of the circles to be resolved, except those with a c.t.c. separation of 240 nm, which is already $\approx$27$\%$ below the Abb{\'e} diffraction limit. Circles with c.t.c. separation of 390 nm and 290 nm have a d.p.v. of $35\%$ and $22\%$ respectively. A halo around the circles is visible and attributed to the sidelobe illumination. However, experimentally (Fig.\ref{fig:2d}\textcolor{urlblue}{(c)}) only the circles in the extreme ends appear separated, and those with c.t.c. separations of 240 nm and 290 nm are merged. We expect the effects of the phase to explain this deviation from numerical simulations (see \textcolor{urlblue}{Supplement} for elaboration). This effect becomes more pronounced in imaging a 2D array of squares as seen in Panel B of Fig.\ref{fig:2d}. There is a marked contrast between numerical simulations (Fig.\ref{fig:2d}\textcolor{urlblue}{(e)}) and experimental results (Fig.\ref{fig:2d}\textcolor{urlblue}{(f)}). The simulations suggest a moir{\'e}-like pattern probably due to interference from sidelobe illumination and reports an average d.p.v of $23\%$ in the middle and $<5\%$ d.p.v in the edges (see \textcolor{urlblue}{Supplementary Figure S10} for line profile). Experimental results are blurry, and the squares are surely not resolved. Also, scan lines are visible in the upper region. \par

It is clear from the above analysis that due to sidelobe illumination only isolated objects can be imaged with super-resolution using the SOL nanoscope. An example application is the resolution of binary stars \cite{Puschmann2005}. \par 
\section{Conclusions}

In summary, we present a novel reflection confocal nanoscope employing a binary metal mask SOL. We relay the SOL intensity pattern onto the object enabling reflection nanoscopy with a super-oscillatory illumination. We experimentally demonstrate the super-resolution capabilities by imaging double bars with a c.t.c. separation of 330 nm using a 632.8 nm excitation wavelength and a 0.95 NA objective. While the SOL nanoscope provides enhanced contrast compared to a LSCM, the sidelobe illumination degrades the imaging properties in the case of complex/large objects. Therefore, the resolution capability of the SOL nanoscope is not an intrinsic property but it very much depends on the characteristics of the scene to be imaged. 
\par

\section*{Acknowledgments}
The authors would like to thank Dr. Benjamin Brenny, Ir. Jos Groote Schaarsberg for discussions and comments on the manuscript, and Dr. Aliasghar Keyvani for help with the AFM. This work was funded by TU/e through the MELISSA PhD project, and by TNO through the Early Research Program (ERP) on "3D Nanomanufacturing".
\newline
\newline
%%%%%%%%%%%%%%%%%%%%%%% References %%%%%%%%%%%%%%%%%%%%%%%%%
See \textcolor{urlblue}{Supplement} for supporting content.
%\bibliography{SOL}
%\end{document}

\clearpage
\setcounter{section}{0}
\setcounter{figure}{0}
%\begin{multicols}[2]
%\documentclass[9pt,twocolumn,twoside]{optica-suppl-materials}
%\setboolean{shortarticle}{false}
%\setboolean{displaycopyright}{false}
%\dates{}
%\doi{\url{http://dx.doi.org/10.1364/optica.XX.XXXXXX}}
%\doi{}
%\title{Reflection confocal nanoscopy using a super-oscillatory lens: supplementary material}

%\author[1, 2, *]{Arvind Nagarajan}
%\author[1, 2]{L. Pjotr Stoevelaar}
%\author[2]{Fabrizio Silvestri}
%\author[3, 4]{Marijn Siemons}
%\author[5]{Venu Gopal Achanta}
%\author[2]{Stefan M. B. B\"{a}umer}
%\author[1, 2]{Giampiero Gerini}

%\affil[1]{Electromagnetics Group, Technische Universiteit Eindhoven (TU/e), 5600 MB Eindhoven, The Netherlands}
%\affil[2]{Optics Department, Netherlands Organization for Applied Scientific Research (TNO),\newline Stieltjesweg 1, 2628 CK Delft, The Netherlands}
%\affil[3]{Department of Imaging Physics, Delft University of Technology, 2628 CJ Delft, The Netherlands}
%\affil[4]{Currently with Cell Biology, Department of Biology, Faculty of Science, Utrecht University, 3584 CH Utrecht, The Netherlands}
%\affil[5]{FOTON Laboratory, Tata Institute of Fundamental Research, Homi Bhabha Road, Mumbai 400005, India}
%\affil[*]{arvind.nagarajan@tno.nl}

% To be edited by editor
% \dates{Compiled \today}
% To be edited by editor
% \doi{\url{http://dx.doi.org/10.1364/optica.99.099999.s001} [supplementary document doi]}
\title{Reflection confocal nanoscopy using a super-oscillatory lens: supplementary material}

%\begin{abstract*}
 %This document provides supplementary information to "Reflection confocal nanoscopy using a super-oscillatory lens". 
%\end{abstract*}
%\setboolean{displaycopyright}{false} %copyright not used in supplementary materials

%\begin{document}
%\maketitle

\section{Device Fabrication}

\begin{figure}[htbp!]
\centering
\includegraphics[width=8cm]{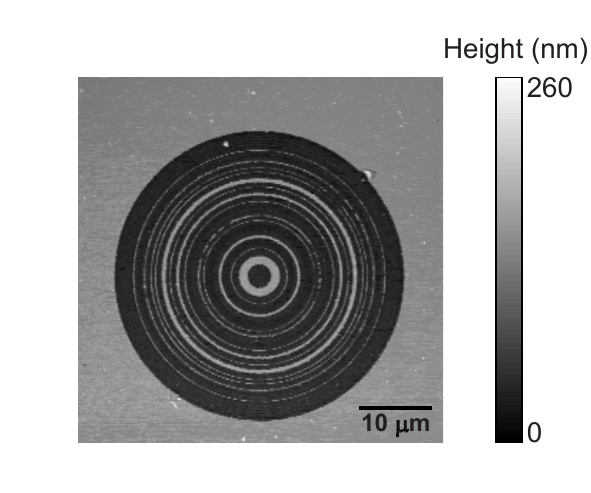}
\caption{AFM scan of the fabricated SOL containing concentric \textit{Ti} rings of 100 nm thickness in a 700 $\mu m$ glass substrate.}
\label{fig:afm}
\end{figure}

The SOL and the object were both fabricated using electron beam lithography. A 700 $\mu m$ thick glass substrate was used to fabricate the SOL. A 100 nm thick \textit{Ti} layer was deposited on the glass substrate by electron beam evaporation. A 200 nm thick poly-(methyl methacrylate) (PMMA) 459 A4 resist layer was then spin coated. Concentric rings (see \cite{Rogers2012a} for dimensions) were exposed on the resist by an electron beam with 20 kV accelerating voltage and a 10 $\mu m$ focusing aperture. The exposed resist was developed in a 1:3 solution of methyl isobutyl ketone (MIBK): isopropyl alcohol (IPA) for 90 s and rinsed in IPA for 60 s. A very thin ($\approx$20 nm) \textit{Al} layer was then deposited on the patterned resist. The pattern was then transferred to the \textit{Ti} layer by reactive ion etching (till the glass substrate) using $SF_{6}$. Here, the \textit{Al} layer serves as an etch mask. The resist was later removed by acetone lift-off, and \textit{Al} was removed by mixture of $H_{3}PO_{4}/HNO_{3}/HAc/H_{2}O$  (TechniEtch Al80, MicroChemicals). The object was fabricated on a $1.1mm$ thick ITO coated glass substrate (Sigma-Aldrich). The procedure described above was also used to expose various test structures in the resist. 100 nm \textit{Au} layer was then sputtered on the patterned resist, and the resist was removed by acetone lift-off. Fig.\ref{fig:afm} shows the Atomic Force Microscope (AFM) scan of the fabricated SOL (poor contrast in scanning electron microscope (SEM) due to charge accumulation in the glass substrate). SEM images of the resulting test structures are shown in the main text.

\section{Necessity for high NA objectives}
\begin{figure}[htbp!]
\centering
\includegraphics[width=8cm]{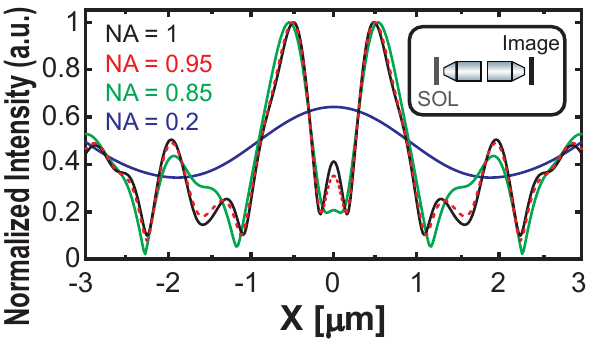}
\caption{Numerically simulated intensity pattern for an imaging system consisting of a SOL and two objectives as shown in the insert. The black curve represents ideal NA=1 system. The red-dotted curve corresponding for NA=0.95 traces the black curve quite well preserving the central sub-diffracted spot. The green curve represents NA=0.85, depicting the onset of this central spot. Super-oscillation is completely destroyed when 0.2 NA objectives are used as seen in the blue curve.}
\label{fig:NA}
\end{figure}

We numerically simulated an imaging system consisting of a SOL and two objectives as seen in the insert of Fig.\ref{fig:NA}  to demonstrate the necessity for high NA objectives. The dimensions of the SOL were taken from \cite{Rogers2012a} and Angular Spectrum Method \cite{Silvestri2017} was used to compute its intensity pattern. Linear systems theory \cite{Goodman1996} is then used to obtain the image.  The simulations were performed for four sets of NA and a line profile along one axis is shown in Fig.\ref{fig:NA}. The black curve represents the ideal NA=1 system where the FWHM of the central hotspot is $180 \pm 20$nm. The red dotted curve represents NA=0.95, and as seen in the figure it traces the black curve quite well preserving the central hotspot. The FWHM is also comparable within the error limits ($170 \pm 20$ nm). Although the intensity of the central hotspot decreases with the NA, it is sub-diffracted till NA=0.85 as seen in the green curve. When we further reduce the NA of the objectives down to 0.2, the superoscillation is completely destroyed as seen in the blue curve. The FWHM of the central spot is now $2\mu$m. Hence, high NA objectives are necessary for capturing the spatial frequency contents crucial for super-oscillation. 

\section{Optimal scan step size}

\begin{figure}[htbp!]
\centering
\includegraphics[width=8cm]{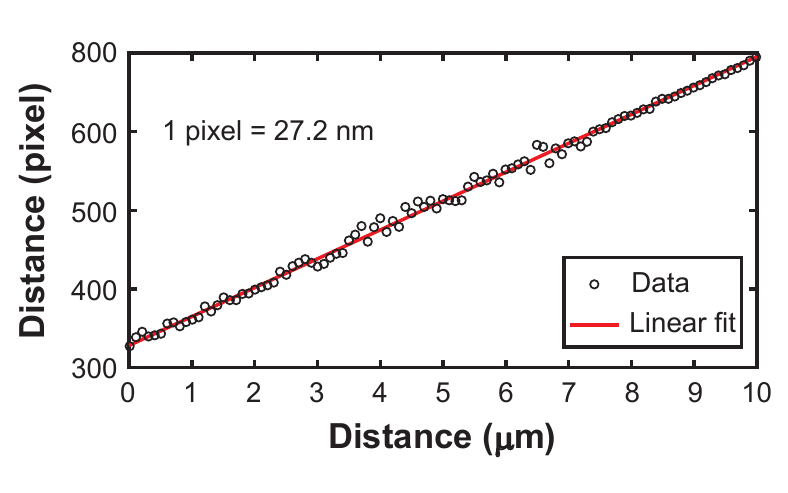}
\caption{Calibrating the pixel size of the nanoscope. Black circles represents the data points and the red line is the linear fit.}
\label{fig:Pixel}
\end{figure}

The SOL is moved over 10 $\mu$m in steps of 100 nm along the vertical direction (Y)(see main text \textbf{Fig.1}). By tracking the position of the central spot on the CCD during this displacement, the effective magnification between the image and the object plane is found as reported in  Figure \ref{fig:Pixel}. The effective pixel size is the slope of the linear fit. The step size of the snake scanning is hence fixed at 30 nm such that the changes between successive scan steps are detectable and not half-pixels. 

\section{The hotspot used for imaging}
\begin{figure}[htbp!]
\centering
\includegraphics[width=8cm]{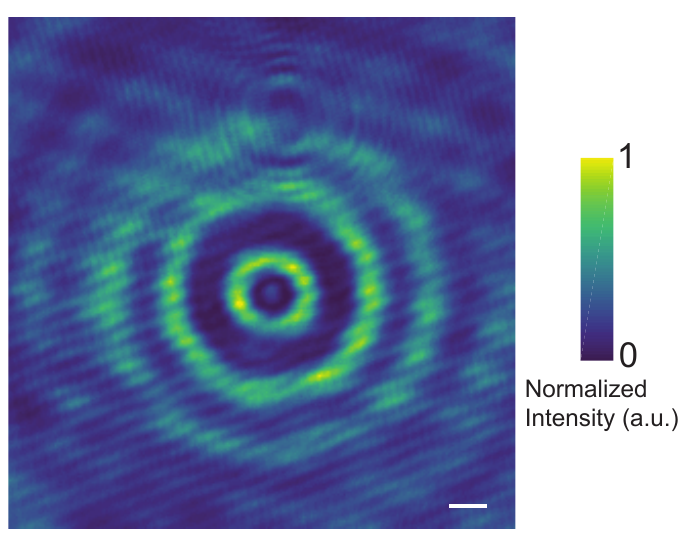}
\caption{The hotspot at Z$\approx$4 $\mu$m. The white scale bar represents 500 nm.}
\label{fig:hotspot}
\end{figure}

\begin{figure}[htbp!]
\centering
\includegraphics[width=8cm]{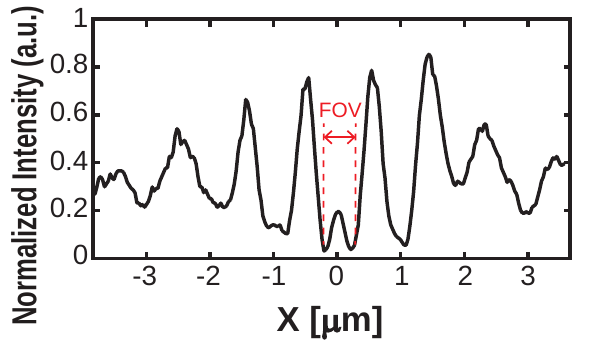}
\caption{Line profile along horizontal axis of the hotspot at Z$\approx 4\mu m$ from the SOL surface. Dotted red lines represents the Field of View (FOV) of the SOL which is $\approx$500 nm.}
\label{fig:hotspot_line}
\end{figure}

The intensity pattern of the hotspot at Z$\approx$4 $\mu$m along with the first few sidelobes is shown in Figure \ref{fig:hotspot}. The side bands are reasonably rotationally symmetric but contains some fringes, which are speculated to be due to ghost reflections \cite{Murray1949}. The intensity of the central spot is $\approx$1/3rd that of the first sidelobe (average over the ring). As seen, the normalized intensity along the first sidelobe is not uniform and contains multiple peaks and valleys (0.75 $\pm$ 0.20 a.u.). Fig.\ref{fig:hotspot_line} shows the line profile across the center of the hotspot. The region within the first sidelobe, containing the central sub-diffracted spot is the Field of View ($\approx$500 nm). 

\newpage
\section{Phase effects}

\begin{figure}[htbp!]
    \centering
    \includegraphics[width=8cm]{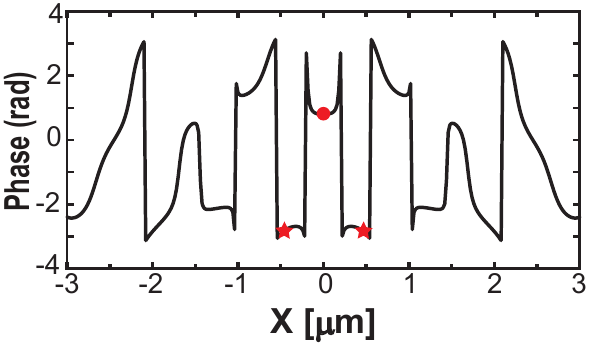}
    \caption{Numerically simulated phase profile of the SOL along one axis. The red dot represents the peak location of the central hotspot and the stars represent the peak locations of the first sidelobe.}
    \label{fig:phase}
\end{figure}

The SOL intensity profile such as the one seen in Fig.\ref{fig:NA} consists of a superimposition of many oscillating functions. Consequently, one can expect the phase to have multiple zero crossings. We obtain the phase profile of the SOL by Angular Spectrum Method \cite{Silvestri2017}. Fig.\ref{fig:phase} plots the line profile of the wrapped phase along one axis of the SOL. Here, the red circle represents the peak location of the central hotspot while the red stars represent the peak locations of the first sidelobe.  The central spot has a different phase compared to the first sidelobe and there is a zero crossing between the two. Therefore, the illumination in the vicinity of sidelobe can induce the reflected light to interfere with itself and create coherent light interference not foreseen by the numerical simulations performed with constant phase CSF. 

\section{Line profiles}

\begin{figure}[htbp!]
\centering
\includegraphics[width=7.5cm]{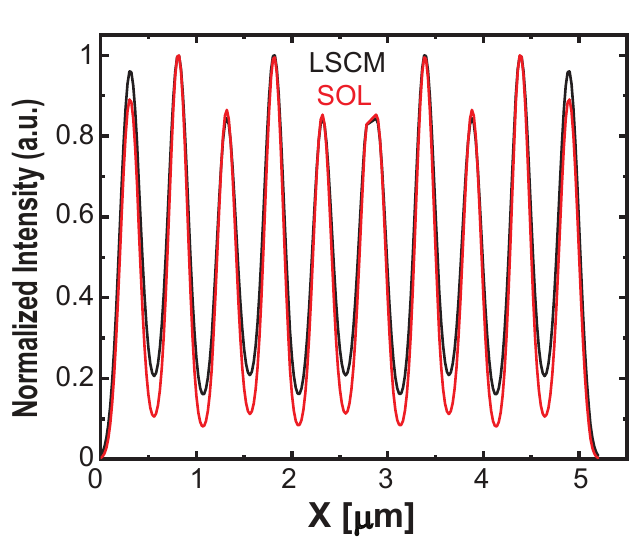}
\caption{Line profile across the center of the Fig.3.(b, c) of the main text. The black line represents the LSCM while the red line represents the SOL Nanoscope.}
\label{fig:1DAS}
\end{figure}

Figure \ref{fig:1DAS} shows the line profiles of numerical simulations on the SOL nanoscope (red line) and the LSCM (black line) in imaging a 1D array consisting of 10 bars with a c.t.c separation of 500 nm. Although both curves overlap with each other, the contrast is better with the SOL nanoscope. 

\begin{figure}[htbp!]
\centering
\includegraphics[width=8cm]{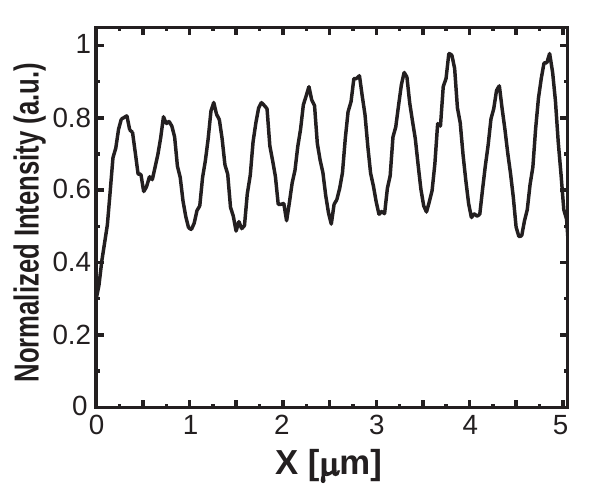}
\caption{Line profile across the center of the Fig.3.(d) of the main text.}
\label{fig:1DA}
\end{figure}

\begin{figure}[htbp!]
\centering
\includegraphics[width=8cm]{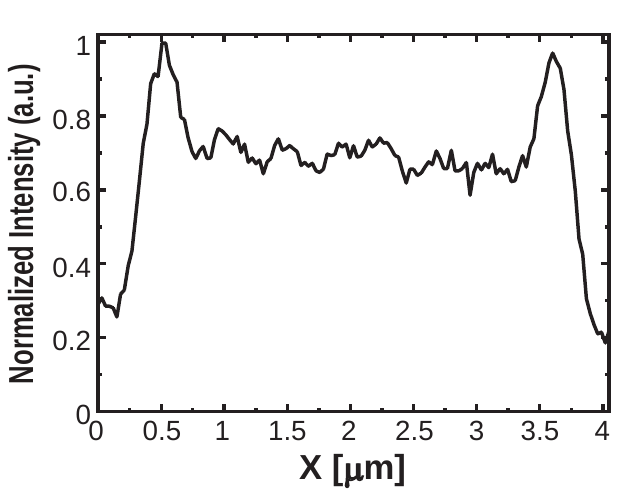}
\caption{Line profile across the center of the Fig.3.(h) of the main text.}
\label{fig:1DB}
\end{figure}

Figures \ref{fig:1DA} and \ref{fig:1DB} show the line profile across the center of the experimentally obtained SOL nanoscope images for a 1D array with a c.t.c spacing of 500 nm and 330 nm respectively. The 1D array with a c.t.c separation of 500 nm is resolved with an average d.p.v of $34.42\%$, and the measured peak to peak separation is 510 $\pm$ 30 nm. The 1D array with a c.t.c. separation of 330 nm is not resolved with the SOL nanoscope. As seen in Figure \ref{fig:1DB}, only the bars in the end are well separated with a d.p.v of $30\%$, and all other bars are not distinguishable. 

\begin{figure}[htbp!]
\centering
\includegraphics[width=8cm]{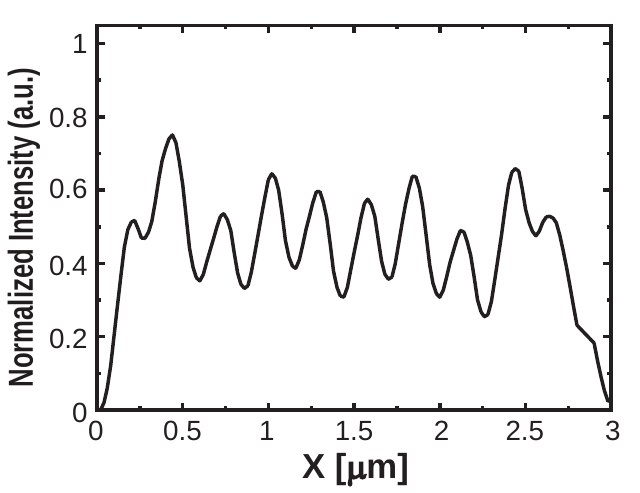}
\caption{Line profile across the center of the Fig.4.(e) of the main text.}
\label{fig:2DS}
\end{figure}

Figure \ref{fig:2DS} shows the line profile across the center of the numerical simulation of the SOL nanoscope in imaging a 2D array of consisting of 10$\times$10 array of squares of size 100 nm in a square lattice with periodicity of 280 nm. 8 peaks in the center are well separated and the peaks in the edges are barely distinguishable. The average peak to peak separation in the center is 280 nm with a d.p.v of $23\%$, while that at the edges is 200 nm with $\approx5\%$ d.p.v. 

% Bibliography
%\bibliography{SOL}

%\end{document}
%\end{multicols}
\newpage
\bibliography{SOL}
\end{document}